\documentclass[useAMS,usenatbib,usegraphicx]{mn2e}

\title[Debris around thick disc stars]
{Forming the first planetary systems: debris around Galactic thick 
disc stars }
\author[C. K. W. Sheehan et al.]{C. K. W. Sheehan$^1$, J. S.
Greaves$^1$\thanks{E-mail: jsg5 at st-andrews.ac.uk}, G. Bryden$^2$, 
G. H. Rieke$^3$, K.Y.L. Su$^3$,
\newauthor M. C. Wyatt$^4$ \& C. A. Beichman$^5$, \\
$^1$SUPA, School of Physics and Astronomy, University of St. Andrews, 
North Haugh, St. Andrews KY16 9SS, UK \\
$^2$Jet Propulsion Laboratory, 4800 Oak Grove Drive, Pasadena, CA 91109, 
USA \\
$^3$Steward Observatory, University of Arizona, 933 North Cherry Avenue, 
Tucson, AZ 85721, USA \\
$^4$Institute of Astronomy, University of Cambridge, Cambridge CB3 0HA, 
UK \\
$^5$Michelson Science Center, California Institute of Technology, 770 
South Wilson Avenue, Pasadena, CA 91125, USA 
}

\begin{document}

\date{Accepted 2010. Received 2010; in original form 2010}

\pagerange{\pageref{firstpage}--\pageref{lastpage}} \pubyear{2010}

\maketitle

\label{firstpage}

\begin{abstract}

The thick disc contains stars formed within the first Gyr of Galactic 
history, and little is known about their planetary systems. The 
Spitzer MIPS instrument was used to search 11 of the closest of these 
old low-metal stars for circumstellar debris, as a signpost that 
bodies at least as large as planetesimals were formed. A total of 22 
thick disc stars has now been observed, after including archival 
data, but dust is not found in any of the systems. The data rule out 
a high incidence of debris among star systems from early in the 
Galaxy's formation. However, some stars of this very old population 
do host giant planets, at possibly more than the general incidence 
among low-metal Sun-like stars. As the Solar System contains gas 
giants but little cometary dust, the thick disc could host analogue 
systems that formed many Gyr before the Sun.
 
\end{abstract}


\section{Introduction}

Debris discs are generated by collisions among comets or asteroids, 
and so are a signpost of planetary systems, indicating that bodies of 
at least planetesimal size formed. The debris absorbs starlight and 
re-emits it at longer wavelengths, and observations since 2003 with 
the Spitzer space telescope have yielded many new debris detections 
in the mid- to far-infrared. The relation between giant planets and 
planetesimal belts is still uncertain. \citet{bryden09} find that the 
incidence and brightness of cool outer-system debris is not affected 
by the presence or absence of gas giants in the system, which is not 
surprising if the formation and evolution of giant planets at a few 
AU and comets at tens of AU are decoupled processes. More subtle 
effects may occur, such as the clearing out of comet belts when giant 
planets migrate, analagous to the Late Heavy Bombardment (LHB) era in 
Solar System history. This could result in brief dusty episodes as 
more planetesimals collide, and subsequent lower dust levels due to 
the ejection of many comets from the system. However, such LHB events 
seem to be atypical, as Spitzer has commonly found old star systems 
that are still very dusty \citep{booth}.

\citet{greaves07} argued that the incidence of giant planet and 
debris systems, and the weak overlap between them, could be explained 
if the mass of refractory elements at the planet-formation stage 
controlled the outcome. Young circumstellar discs with a large 
`solid' mass could quickly form planet cores that go on to accrete 
gas atmospheres, while less substantial discs only form planetesimals 
by the time the gas disperses and so evolve into debris systems. This 
model (see also \citet{wyatt07}) addresses the differing metallicity 
dependences observed -- metal-rich discs tend to have a higher solid 
mass and so a greater probability for building a planet core, while 
lower solid reservoirs are needed to build comets and so debris has 
little dependence on metallicity 
\citep{trilling,greaves07,bryden06,beichman}.

\begin{table}
 \caption{Properties of the thick disc stars. The first two sets of 
stars form the volume-limited sample and the third set includes those
outside the complete volume or more recently identified as thick disc
objects (see text). Metallicities are for Fe except HD~19034 and 
101259 which are combined metals. The last columns are the observed 
date and number of exposures for the new data (see text). 
}
 \label{tab:list}
 \begin{tabular}{@{}rcccccc}
  \hline
HD 	& [Fe/H]& d	& type	& observation	& $N_{exp}$	\\
	&	& (pc)	&	& date		& (24,70 )  	\\
  \hline
\multicolumn{6}{c}{new observations} \\
19034	& -0.41 & 35    & G5		& 27/08/07	& 2,20	\\
22879	& -0.84 & 24    & F9V		& 03/03/07	& 2,5	\\
29587	& -0.61 & 28    & G2V		& 29/10/07	& 2,10	\\
30649	& -0.50 & 30    & G1V-VI	& 29/10/07	& 2,15	\\
65583	& -0.67 & 17    & G8V		& 07/11/06	& 2,5	\\
68017	& -0.42 & 22    & G4V		& 07/11/06	& 2,5	\\
88725	& -0.70 & 36    & G1V		& 29/05/07	& 2,10	\\
111515	& -0.52 & 33	& G8V		& 12/07/07	& 2,20	\\
144579	& -0.70 & 14    & G8V		& 13/04/07	& 2,5	\\
218209	& -0.43 & 30    & G6V		& 16/08/06	& 2,20	\\
221830	& -0.45	& 32	& F9V		& 21/08/07	& 2,10	\\
\multicolumn{6}{c}{other volume-limited objects} \\
3795	& -0.59 & 29    & K0V		& 		& 	\\
63077	& -0.83 & 15    & F9V		& 		& 	\\
114729	& -0.25 & 35    & G0V		& 		& 	\\
114762	& -0.70 & 41	& F9V		& 		& 	\\
\multicolumn{6}{c}{other thick disc stars} \\
6582	& -0.86 & 8	& G5Vb		& 		&	\\
23439	& -1.03	& 24	& K2V,K3V	&               &       \\
76932	& -0.86 & 21	& F9V		&               &       \\
101259	& -0.23 & 65	& G8IV		&               &       \\
106516	& -0.74 & 23	& F5V		&               &       \\
157214	& -0.41 & 14	& G0V		&               &       \\
  \hline
 \end{tabular}
\end{table}

Here we investigate the formation of planetary systems by observing 
some of the oldest Galactic stars with Spitzer. A good test of models 
is to observe old, metal-poor stars. If these proceeded only 
as far as planetesimal growth, a high incidence of debris might be 
seen, but perhaps modified by the longer time available for 
collisional grinding and dust removal compared to mid-Galactic-age 
stars like the Sun \citep{wyatt08} . The sample of stars studied here 
is identified as nearby members of the thick disc -- stars that 
formed within about the first Galactic Gyr \citep{chiappini} and have 
orbits inclined with respect to the Galactic Plane. These systems are 
about twice as old as the Solar System.

\section{Sample selection}

The targets (Table~1) were initially derived from \citet{karatas}, 
who identified 22 Galactic thick disc stars near the Sun. The 
identification uses criteria of low metallicity and rotation speed, 
with values centred around [Fe/H]~=~--0.5 and 180~km/s respectively, 
removing halo stars that have lower $v_{rot}$ for the same 
metallicity. The sample was then reduced to the 15 stars that lie 
within 40~pc of the Sun, to make the best use of the sensitivity of 
the MIPS instrument \citep{rieke04}. Four of these stars had been 
observed in previous Spitzer programs 
\citep{beichman,trilling,bryden09}, and the other 11 were observed 
for our Spitzer program number 30339. Since the survey was proposed, 
\citet{reddy} identified further thick disc stars, several of which 
were observed in various Spitzer programs, as was one of the more 
distant stars of \citet{karatas}, HD~101259. The total number of stars 
discussed here is 22, of which 19 have data at two Spitzer 
wavelengths.


\begin{table}
\caption{Observed fluxes F and errors, and excesses E,R derived from 
observed fluxes compared to the photospheres. E[24] is expressed in 
magnitudes while R70 is the ratio of the observed and photospheric 
fluxes, with a significance for any dust of $\chi_{70}$ i.e. excess 
flux above the photosphere divided by its error.
}
 \label{tab:list}
 \begin{tabular}{@{}rccrrrr}
  \hline
HD	& F24		& F70		& E24		& R70	& $\chi_{70}$ 	\\
	& (mJy)		& (mJy)		& (mag)		& 	& 		\\
  \hline
3795	&137.7 $\pm$0.7	&19.5 $\pm$2.7	&0.03		&1.3 	&1.5	\\
6582	&350   $\pm$1	&		&0.05		&	&	\\
19034	&19.9  $\pm$0.5	&0.1  $\pm$2.4	&-0.02		&0.0	&-0.9	\\
22879	&59    $\pm$0.4	&15.4 $\pm$5.2	&-0.04		&2.4	&1.7	\\
23439A	&25.3  $\pm$0.5	&		&-0.01		&	&	\\
23439B	&19.9  $\pm$0.5	&		&-0.05		&	&	\\
29587	&38.8  $\pm$1.2	&2.2  $\pm$3.9	&-0.02		&0.5	&-0.5	\\
30649	&48.2  $\pm$1.2	&1    $\pm$3.6	&0.01		&0.2	&-1.2	\\
63077	&227.8 $\pm$1.1	&15.5 $\pm$4.1	&0.00		&0.6	&-2.3	\\
65583	&65    $\pm$0.3	&12.2 $\pm$6.5	&-0.01		&1.7	&0.8	\\
68017	&64.8  $\pm$0.3	&6.2  $\pm$6.1	&-0.02		&0.9	&-0.2	\\
76932	&133.8 $\pm$0.7	&14.7 $\pm$2.1	&0.03		&1.0	&-0.0	\\	
88725	&24.3  $\pm$0.4	&2.3  $\pm$3.2	&-0.03		&0.9	&-0.1	\\
101259	&152.3 $\pm$0.7	&23.3 $\pm$2.2	&0.06		&1.4	&2.6	\\
106516	&80.4  $\pm$0.4	&6.2  $\pm$6.2	&-0.04		&0.7	&-0.4	\\
111515	&20.2  $\pm$0.5	&0.9  $\pm$2.3	&-0.02		&0.4	&-0.6	\\
114729	&62.1  $\pm$0.7	&7.9  $\pm$2.2	&-0.02		&1.2	&0.5	\\
114762	&32.4  $\pm$0.3	&0.2  $\pm$3.1	&-0.05		&0.1	&-1.1	\\
144579	&88.6  $\pm$0.5	&13.1 $\pm$4.9	&-0.02		&1.3	&0.7	\\
157214	&217.5 $\pm$1	&23.9 $\pm$4.2	&0.04		&1.0	&-0.0	\\
218209	&33.4  $\pm$1.4	&0.8  $\pm$4.2	&-0.03		&0.2	&-0.7	\\
221830	&53.5  $\pm$0.4	&9.6  $\pm$2.7	&-0.02		&1.6	&1.4	\\
  \hline
 \end{tabular}
\end{table}

The (logarithmic) metallicities in this sample are -0.23 to -1.03, 
considerably lower than the mean of --0.06 with standard deviation of 
0.25 for local thin disc stars \citep{valenti}. For nine of the 
volume-limited objects, \citet{takeda} have calculated the ages based 
on isochrone fitting, a method that is most accurate for old stars 
that have evolved off the main sequence. Where the age is well 
determined (uncertainties of $\la 30$~\% for HD~3795, 30649, 68017, 
114729, 218209 and 221830), these stars are dated at 9.6--11.6 Gyr 
old. This is in very good agreement with their thick disc membership 
\citep{karatas}, and independent age estimates are consistent with 
this conclusion. We checked for chromospheric activity, and although 
current calibrations of age versus activity \citep[e.g.]{mamajek} are 
only for roughly solar metallicity, all of our program stars are 
inferred to have low activity consistent with ages greater than 
4~Gyr.

\section{Observations and data reduction}

The eleven new stars were observed between 2006 August and 2007 
October, using MIPS at 24 and 70~$\umu$m (Table~2). Exposure times 
were 3 and 10 seconds at these wavelengths respectively, multiplied 
by the number of repeats given in Table~1. The data reduction and 
instrument calibration procedures are described in 
\citet{engelbracht,gordon}. The initial data reduction was by 
aperture photometry as described in \citet{trilling}. Subsequently, 
these observations plus the archival data for the other stars were 
re-reduced homogeneously as part of a Spitzer legacy catalogue 
\citep{su} and these values are presented.


\begin{figure}
\label{fig1}
\includegraphics[height=82mm,angle=270]{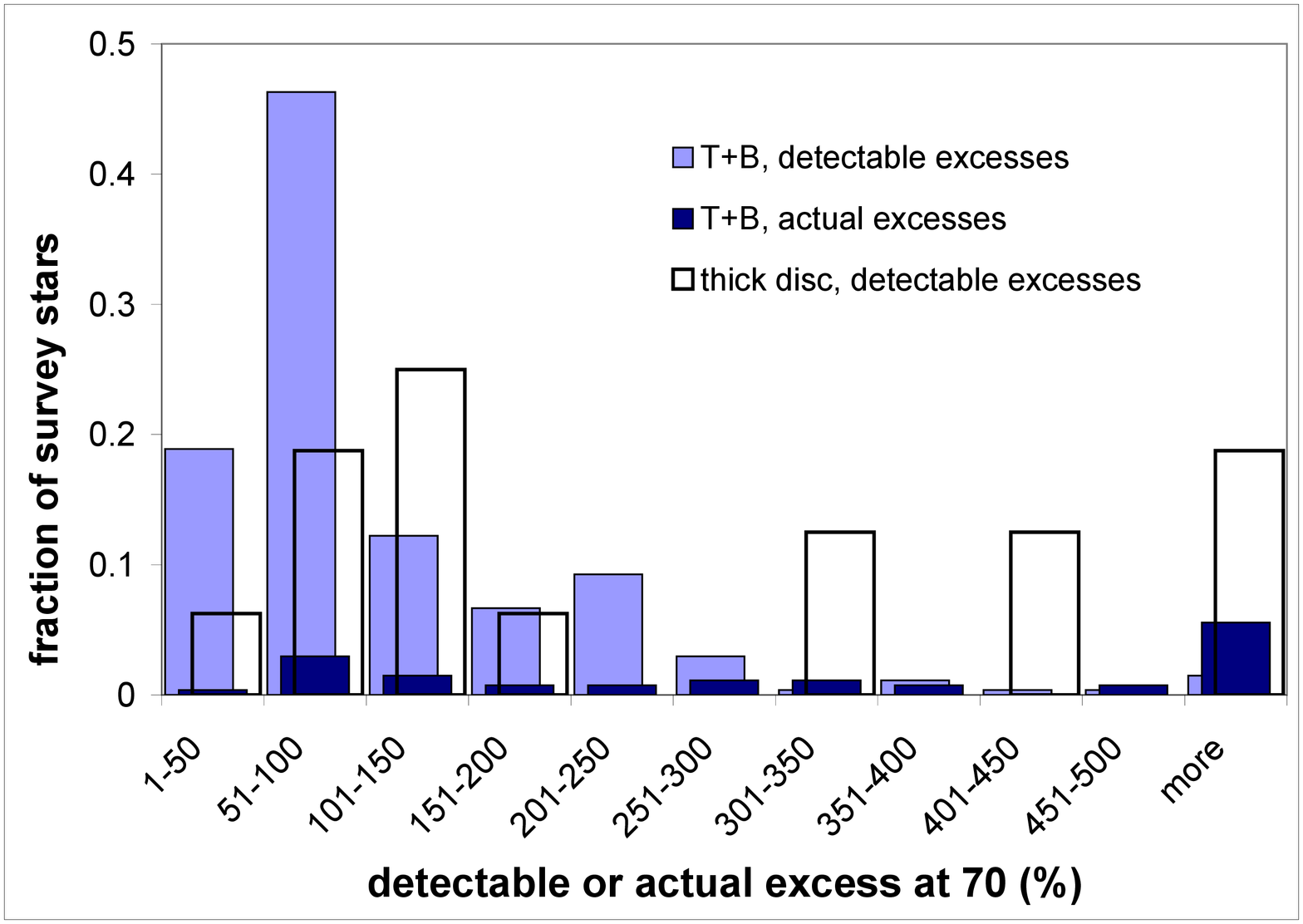}
\caption{Plot of survey sensitivity at 70~$\umu$m (Table~2) compared 
to Sun-like stars observed by \citet{trilling,beichman}, denoted by 
T+B. The bins show relative numbers of survey stars where, if dust was 
present at that excess level, it could be detected with $3\sigma$ 
confidence. The actual distribution of T+B dust levels is also 
plotted (dark blue bars). The excesses are above the photosphere, 
i.e. R70-1. 
}
\end{figure}

To identify candidate excesses, photospheric signals were predicted 
systematically. K-band magnitudes from 2MASS were extrapolated to 
24~$\umu$m based on a set of photospheric solar-type stars. A few 
stars have K-band magnitudes $\la 4.5$ that would saturate the 2MASS 
images, and for these heritage photometry was transformed into the 
2MASS system. These fluxes, for HD 3795, 6582, 63077, 76932 and 
157214, will be discussed in the legacy papers (Su, Rieke \& 
Whitelock, in prep.); the last value is rather uncertain with a rms 
error of $\sim 0.06$~mag. As no 24~$\umu$m excesses were found, the 
excess at 70~$\umu$m was then calculated relative to the 24~$\umu$m 
flux. The significance of excess at 70~$\umu$m includes 5\% 
photometric errors plus the statistical error.

One system, HD23439, is a binary with two resolved components that 
are treated separately here. Amongst the rest of the systems, HD 
221830 has a binary companion at approximately 9~arcsec separation 
that could add 10-15~\% extra photospheric flux at 70um (based on the 
relative K-band brightnesses), and HD 111515 has a similarly 
separated companion of a fainter nature, but both of these 
contaminants are neglected here.

\begin{figure}
\label{fig2}
\includegraphics[width=85mm,angle=0]{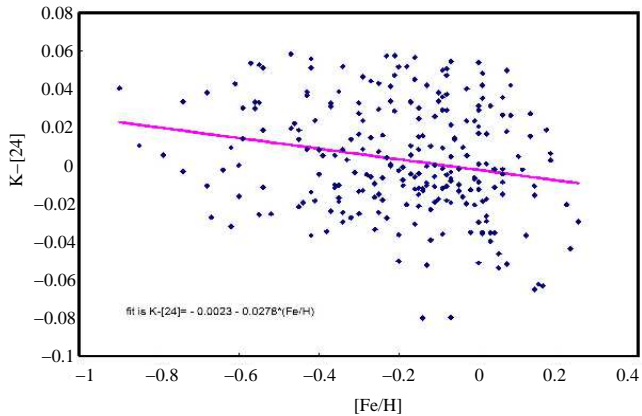}
\caption{K-[24] colour against [Fe/H] (from NSTARS, at  
http://nstars.nau.edu) and linear fit. Data are from various Spitzer 
surveys, re-reduced homogeneously.
}
\end{figure}

\section{Results}

\subsection{Survey limits}

From Table 2, we find no dust excess towards any of the stars. At 
24~$\umu$m, the differences between observed and photospheric fluxes 
vary between -0.05 and 0.06 magnitudes, so there are no positive 
outliers. The standard deviation of the excesses is 0.03 magnitudes 
and the mean is -0.01 magnitudes. At 70~$\umu$m, there is one 
candidate excess of $+2.6\sigma$, but the most negative value is at 
$-2.3\sigma$, so there are again no notably positive outliers. 

About half the stars were not detected at all at 70~$\umu$m. This 
is mainly attributable to larger distances of rare thick disc 
objects compared to thin disc stars. For example, the median 
target distance is 28~pc compared to 20~pc in \citet{trilling}, 
while the average 70~$\umu$m noise level is about 5~mJy in both 
cases. Three of the stars (HD~29587, 30649, 218209) observed here 
also had high sky backgrounds that can increase the noise. 
Figure~1 assesses the sensitivity achieved within the thick disc 
sample. Survey stars are binned here according to the minimum 
excess that could be detected with $3\sigma$ confidence. The 
fraction of thick disc stars where only excesses above three 
times the photosphere can be detected (i.e. bins from 301-350~\% 
upwards) is rather large. In the general stellar surveys, the 
sensitivity attained was generally better than this (light blue 
bins). However, many excesses above 300~\% were seen in those 
studies (dark blue bins), suggesting that our survet could be 
adequately deep if the thick disc stars have typical debris . The 
biases in comparing the incidence of debris for the thin and 
thick disc stars are discussed further below.

It was noted that some stars have 70~$\umu$m fluxes well {\it 
below} the photospheric prediction, although with large errors. 
Since the MIPS calibration is reliable over a wide range of 
fluxes \citep{gordon}, we considered whether the stars could be 
under-bright for reasons related to metallicity. However, for a 
wide sample of stars observed at 24~$\umu$m, there is a mild 
trend of {\it enhanced} flux at low [Fe/H], contributing about 
0.02 magnitudes at our median metallicity of --0.6 (Figure~2). If 
the 24~$\umu$m fluxes are normal or slightly high, it would be 
surprising if metal-poor stars switched to being under-bright 
somewhat longwards\footnote{The stars HD 65583, 68017, 144579 and 
157214 have also been observed with Spitzer IRS and have no 
excesses out to 34~$\umu$m \citep{lawler}.}, at 70~$\umu$m. Since 
stars observed with good signal-to-noise tend not to be 
under-bright, an optimal scaling test was carried out to check 
the observed to predicted flux ratio. The scale factor is found 
to be $1.08 \pm 0.06$, so we conclude that there is no general 
70~$\umu$m flux deficit and the faint results are by chance. 
Also, the distribution of significance values $\chi_{70}$ 
(Table~2) is a slightly broadened Gaussian ($\sigma$~=1.15) 
centred close to zero, indicating the 70~$\umu$m results are not 
anomalous.

\begin{figure}
\label{fig3}
\includegraphics[width=85mm,angle=0]{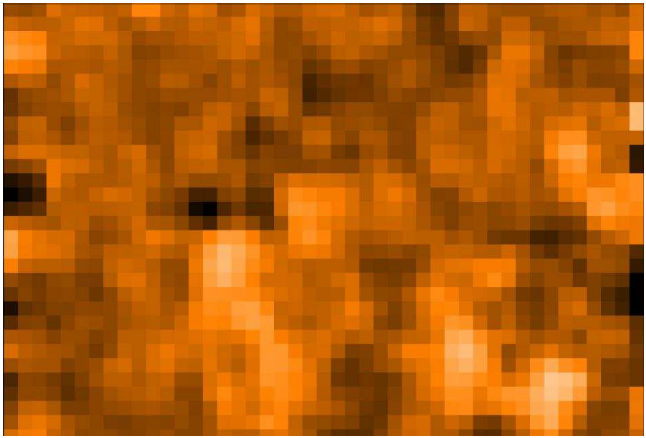}
\caption{Co-added 70 $\umu$m data, showing the flux range 5-95~\%. The 
averaged star signal is visible at the centre.
}
\end{figure}

\subsection{Dust excesses}

To search more deeply for cool excess among the ensemble, frames for 
the 11 new 70~$\umu$m observations were stacked about the stellar 
positions, taking the resulting mean per-pixel signals. In Figure~3, 
the mean photosphere is now faintly visible at centre, although 
background structures from individual frames are also present. 
Aperture photometry gives a mean flux of $4.6 \pm 1.2$~mJy, 
consistent with minimal debris given the mean 70~$\umu$m photospheric 
flux of 5.2~mJy.

A better estimate of the net cool dust within the whole thick disc 
sample can be found from a weighted average of the 70~$\umu$m 
excesses (Table~2). We define E70~=~R70-1 and weight the values by 
1/error$^2$, where the error in E70 can be written as E70/$\chi$70. 
The net excess for the 19 stars with 70~$\umu$m data is $0.07 \pm 
0.34$, i.e. $\la 40$~\% above the photosphere at the 1-sigma upper 
bound. This result is however dominated by five stars with low-noise 
data. For a stellar effective temperature of around 5600~K and dust 
peaking in brightness at 70~$\umu$m, i.e. of temperature around 50~K, 
an excess of $< 40$~\% gives an upper limit on fractional dust 
luminosity $L_{dust}/L_{star}$ of $4\times 10^{-6}$ \citep{beichman}.

Approximately 80~\% of nearby Sun-like stars are also dust-free at 
this level individually \citep{gw10}. Hence, if the thin and thick 
disc populations have similar debris incidence and levels, the null 
result here for only 19 stars is not surprising. In more detail, the 
expected number of debris detections at 70~$\umu$m can be estimated 
from Figure~1, taking into account the limiting depths of the thick 
disc data. The number of detection per 50\%-wide excess bin is then 
the fraction in the general (T+B) surveys, scaled to 19 stars, and 
multiplied by the fraction of thick disc targets where such an excess 
was detectable. This yields a cumulative 2.1 expected detections, 
with a Poisson error of $\pm 1.5$, so finding no debris systems is 
within the errors. Similarly, there is a 5~\% incidence of 24~$\umu$m 
excess for Sun-like stars in general \citep{koerner}, so finding none 
among 22 stars here is consistent with similar populations. 
Studying trends by age, \citet{gaspar} found a 24~$\umu$m 
debris-incidence of approximately $4.1 \pm 1.5$~\% at 3-10~Gyr. In 
theoretical models, some slow fading is predicted, with fractional 
dust luminosity decreasing e.g. by 25~\% by 11~Gyr compared to 
typical mid-age Galactic systems at about 5~Gyr \citep{lohne}.

Statistics are ultimately limited by the small numbers of thick disc 
stars actually passing through the Solar vicinity at the present 
time, and in our volume-limited sample we have already searched for 
debris among the nearest 15, or 70~\%, of objects catalogued by 
\citet{karatas}. Fainter levels of debris or cooler dust may be 
accessible to the Herschel satellite, now observing at wavelengths of 
70 to 500~$\umu$m. 

\section{Discussion}

The debris incidence in the thick disc is either low or normal; 
we can rule out a particularly {\it high} incidence of excess in 
these metal-poor systems. Recent theoretical work by 
\citet{johansen09} has suggested that low-metal systems could 
have late planetesimal formation as the gas disc is dispersed, in 
which case more collisional debris might have been expected 
around the thick disc stars, but this is not seen here. In 
contrast, some thick disc stars do host giant planets -- an 
outcome thought to reflect more efficient planetesimal-building 
at early stages. In our sample HD~114729 has a gas giant 
\citep{butler}, and four other good planet-candidates exist for 
thick disc stars \citep{gonzalez}. The M sini of these planets 
range from 0.5 to 7 Jupiter masses and the semi-major axes from 
0.15 to 2~AU.

\begin{figure}
\label{fig4}
\includegraphics[width=70mm,angle=270]{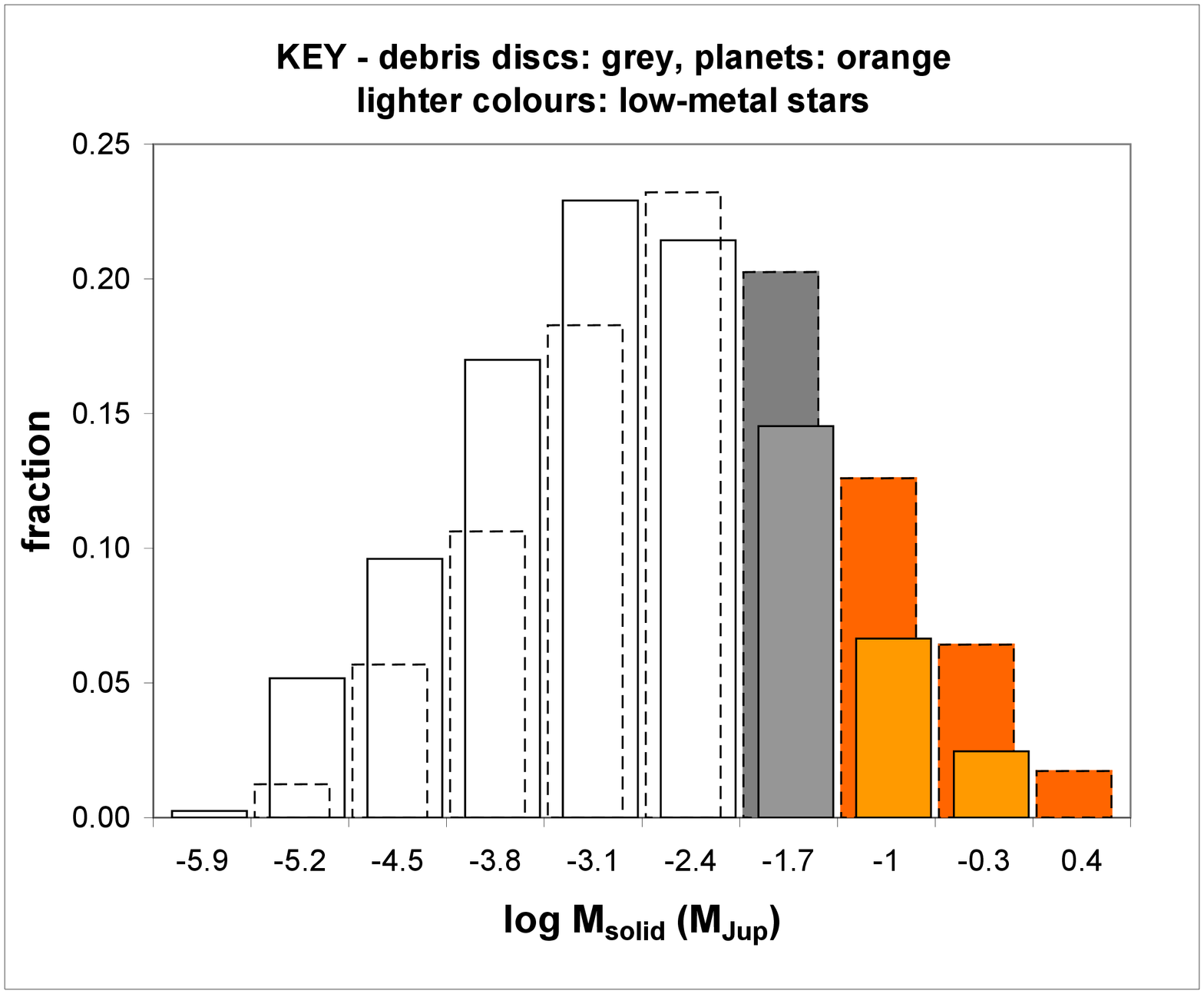}
\caption{Distributions of solid masses available in the discs at early 
times (see text), for thin disc stars and the general nearby star 
population (shown with dashed boxes and darker shading). Giant 
planets can form from the three right-most bins (orange) and debris 
discs from the adjacent bin (grey); mass labels refer to the lowest 
values in the bin.  
}
\end{figure}

Figure~4 shows the predictions of our model \citep{greaves07} based 
on the mass of solids available at the planet(esimal) formation 
stage. The model is based on a threshold of solid material needed for 
a successful outcome such as building a gas giant core, and with the 
solid mass within the disc scaling by the metal fraction in common 
with the star. The $M_{solids}$ distribution is obtained by 
multiplying the log-normal distribution of overall disc masses 
\citep{andrews} by the log-normal distribution [Fe/H] for the 
population of interest, here with mean and standard deviation of -0.6 
and 0.2 respectively. The thresholds for forming detectable debris 
discs and gas giants are around 0.02 and 0.1 Jupiter masses in solids 
respectively \citep{greaves07}. For simplicity we neglect systems 
with both planets and debris, since the two phenomena appear 
uncorrelated \citep{bryden09}, and so this is a small subset of 
Sun-like stars.

Figure~4 then predicts that 15~\% of thick disc stars should form 
giant planets, with a further 9~\% forming comet belts. Around the 
general local population of Sun-like stars, the corresponding numbers 
are 21~\% and 20~\%. The differences are modest because the 
availability of solid material is largely dictated by the total disc 
masses, which have a very broad distribution of $\sigma = 1.15$~dex, 
compared to the standard deviations around 0.2 dex for the 
metallicities. Hence, while the peaks of the solid-mass distributions 
do shift for differing metals (Figure~4), the upper tails associated 
with forming planet(esimals) are broad and do not have greatly 
different numbers of stars.

These values are compatible with survey data, taking into account 
that not all systems of planets and debris have yet been discovered. 
In the general stellar population, the extrapolated planet total is 
17-20~\% \citep{cumming} and the debris disc count at 70~$\umu$m is 
around 14-20~\% \citep{trilling,gw10}, which agree well with the 
predictions of 21 and 20~\% respectively. For the thick disc stars, 
the debris incidence is $\la 5$~\% versus the 9~\% predicted, 
and the planet count is estimated at $\sim 6-11$~\%, versus 15~\% 
predicted. It has been noted that lower-metal discs may disperse 
faster \citep{yasui}, and this could reduce the efficiency of growing 
planet(esimals) compared to our simple model.

The thick disc planet frequency is poorly known, with our 6~\% 
estimate derived from three planetary systems among 52 thick disc 
objects that have [Fe/H]~$< -0.2$ and [$\alpha$/Fe]~$> 0.2$ 
\citep{reid,valenti}, while the 11~\% is for two out of 18 systems in 
the Fe,Mg-selected, volume-limited, thick disc sample of 
\citet{fuhrmann}. These planet frequencies are actually {\it higher} 
than for low-metal stars in general, such as the incidence of 1~\% 
for periods of less than three years found in a recent survey by 
\citet{sozzetti}. Conditions that could promote the formation of more 
gas giants in the old thick disc population compared to younger 
low-metal thin disc stars are at present unclear.

\section{Conclusions}

While the scarcity of nearby thick disc stars makes it difficult to 
study the population, there are examples of disc and planet systems 
around some comparably old stars. A gas giant is known around the 
post main sequence star V391 Peg of age $\ga 10$~Gyr 
\citep{silvotti}; the $\approx 10$~Gyr-old G8 V star $\tau$~Ceti has 
a debris disc \citep{greaves04}; and the oldest known star with both 
a planet and debris is 70~Vir at $\approx 9$~Gyr \citep{bryden09}. 
These discoveries point to some of the first stars in the Galaxy 
having had the potential to form planetary systems akin to those 
around typical stars seen today. If some thick disc stars have 
minimal cool debris but do host gas giants orbiting at a few AU, 
these would be analogues to the Solar System formed many Gyr before 
the Sun.

\section*{Acknowledgments} 

CKWS thanks STFC for a studentship, and JSG thanks STFC for a 
fellowship, in support of this work.

\bsp

\label{lastpage}

\end{document}